\documentstyle[twocolumn,epsfig,color]{mn}
\newcommand{\mincir}{\raise
-2.truept\hbox{\rlap{\hbox{$\sim$}}\raise5.truept
\hbox{$<$}\ }}
\newcommand{\magcir}{\raise
-2.truept\hbox{\rlap{\hbox{$\sim$}}\raise5.truept
\hbox{$>$}\ }}
\newcommand{\minmag}{\raise-2.truept\hbox{\rlap{\hbox{$<$}}\raise
6.truept\hbox{$>$}\ }}
\newcommand{\mnras}{MNRAS}
\newcommand{\apj}{ApJ}
\newcommand{\apjl}{ApJL}
\newcommand{\apjs}{ApJS}
\newcommand{\aj}{AJ}
\newcommand{\aap}{A\&A}

\newcommand{\lcdm}{$\Lambda$CDM}
\newcommand{\rundm}{run$_{DM}$}
\newcommand{\runb}{run$_{baryons}$}

\title[On the coldness of  the local Hubble flow: the role of baryons]{On the coldness of  the local Hubble flow: the role of baryons}

\author[S.~Peirani et al.]
 {S. Peirani$^{1}$\thanks{E-mail: peirani@iap.fr}\\
$^{1}$ Institut d'Astrophysique de Paris, 98 bis Bd Arago, 75014 Paris, France -\\
 Unit\'e mixte de recherche 7095 CNRS - Universit\'e Pierre et Marie Curie. \\
}

\begin{document}

\maketitle

\begin{abstract}
Our aim is to investigate whether the presence of baryons 
     can have any significant influence on  the properties of the
     local Hubble flow which has proved to be ``cold''.
We use two cosmological zoom simulations in the standard $\Lambda$CDM
     cosmology with the same set of initial conditions
     to study the formation of a local group-like system within
     a sphere of $\sim 7\,h^{-1}$ Mpc.
     The first one is a pure dark matter simulation (\rundm)  while
     a complete treatment of
     the physics of baryons is introduced in the second one (\runb). A simple
     algorithm based on particles identity allows us to match haloes from the two runs.
We found that galaxies identified in \runb \, and their corresponding 
     dark matter haloes in \rundm \, have very similar spatial distributions
     and dynamical properties on large scales. Then, when analyzing the velocity
     field and the deviation from a pure Hubble flow
     in both simulations, namely  when computing the dispersion of peculiar velocities  of
     galaxies $\sigma_*(R)$ and those of their corresponding dark matter
     haloes $\sigma_{DM}(R)$ in \rundm,
     we found no particular differences for distances $R=$1 to 8 Mpc
     from the local group mass center. 
     This  suggests that the presence of baryons have
     no noticeable impact on the global dynamical properties of  the local Hubble flow
     within  such distances.
     Then, the results indicate that the ``true'' $\sigma_{*}(R)$ values can be estimated
     from the pure dark matter simulation with a  mean error  
     of  3 km/s  when dark matter haloes are selected 
     with  maximum  circular velocities of $V_c\geq$30 km/s, corresponding  
     to  a population of dark matter haloes in \runb \,
     that host galaxies.
     By investigating  the properties of the Hubble flow at distances
     $R\sim$0.7 to  3 Mpc,
     we also found that the estimation of the total mass enclosed at the radius of the 
     zero-velocity surface $R_0$, using 
     the spherical infall model adapted to $\Lambda$CDM,
     can be  underestimated by at least 50\%.

\end{abstract}


\section{Introduction}

The study of dynamical and photometric properties of galaxies in the local
universe represents an ideal framework to confront predictions of the 
standard $\Lambda$ cold dark matter ($\Lambda$CDM) model with observations.
 In particular, 
an interesting feature of the  dynamical properties of the local universe is
that the Hubble flow is rather ``cold'', e.g. 
the dispersion in the peculiar velocities within the Hubble flow
is quite small, namely $\leq$ 100 km/s (Sandage \& Tammann 1975; Giraud 1986;
Schlegel et al. 1994; Ekholm et al. 2001, Karachentsev et al. 2003; 
Macci{\`o}, Governato \& Horellou 2005; Tikhonov \& Klypin 2009). 

The presence of the dark energy was proposed as a possible explanation
for the smoothness of the local Hubble flow, first argue by Baryshev et al.
(2001), Chernin et al. (2001, 2004, 2007) and Teerikorpi et al. (2005) and supported by
Macci{\`o}, Governato \& Horellou (2005) using a set of N-body simulations. However,
Hoffman et al. (2007) compared results from their own simulations of
CDM and $\Lambda$CDM cosmologies with identical parameters,
apart from the presence or not of the cosmological constant term.
They claimed that no significant differences were noticed
in the velocity flow around galaxies having properties similar to those
observed in the neighborhood of the Milky Way.  A similar conclusion was
obtained more recently by Martinez-Vaquero et al. (2009) using a
range of N-body simulations in different cold dark matter
scenarios. They conclude that  the main dynamical parameter that can affect
the coldness of the flow is the relative isolation of the Local Group.
Another approach was proposed by Peirani \& de Freitas Pacheco (2006, 2008) who derived  
a  velocity-distance relation by modifying the Lema\^itre-Tolman model (TL) 
with the inclusion of a cosmological constant term. They found that this new relation, 
which describes the behavior of the Hubble flow near the central dominant objects ($\leq$2-3 Mpc), 
 in addition to that derived from the
``canonical''  model ($\Omega_m=1$), provide equally acceptable
fits to the existing available data. As a consequence, no robust conclusion about
the effects of the cosmological constant
on the dynamics of groups could be established.
Moreover, Axenides \& Perivolaropoulos (2002) studied the dark energy effects
in the growth of  matter fluctuations in a flat universe. They concluded
that the dark energy can indeed cool
the local Hubble flow but the parameters required for the predicted velocity 
dispersion to match the observed values are
out by observations that constrain either the present
dark energy density or the equation of state parameter w(=$P_x/\varepsilon_x$).

Other cosmological models were proposed to study  the local Hubble flow. 
Dark matter simulations by Governato et al. (1997) for cosmological models 
with $\Omega_m = 1$ or $\Omega_m = 0.3$ are, according to these
authors, unable to produce systems embedded in regions of ``cold'' flows, i.e., with
1-D dispersion velocities of approximately 40-50 km/s. 
Recently, Tikhonov et al. (2009a, 2009b)  have compared the observed spectrum of minivoids
in the local volume with the spectrum of minivoids determined from the
simulations of CDM or WDM models. They found that model predictions  and
observations match very well provided that galaxies can only be hosted by
dark matter haloes with  circular velocities greater than 20 km/s (for WDM)
and 35 km/s (for \lcdm). They have also derived rms deviations from the Hubble
flow which seem to be consistent with observational values.

All those past numerical works have used collisionless N-body simulations to
investigate the puzzle of coldness of the local Hubble flow and thus,  the 
velocity dispersions have been derived by considering the dynamical properties
of dark matter haloes supposed to host galaxies.  
However, it is nonetheless not obvious that
positions and velocities of galaxies and their host dark matter haloes are
identical. A first element of answer was given by Weinberg et al. (2008) 
who have studied  the subhalo and galaxy populations in a galaxy group
simulated either by a collisionless cosmological simulation or a hydrodynamics
simulation with the same initial conditions.
They found that positions and masses of large subhaloes are very 
similar in both runs while they can be different for low mass subhaloes 
whose orbits can be more easily modified by the host halo potential. 
Past works also suggest the existence of bias between the star and dark matter  components
(see for instance, Carlberg, Couchman \& Thomas 1990; Zhao,
Jing \& B\"orner 2002;  Sousbie et al. 2008). 
In the present work, we study  possible effects of the presence of baryons in the
dynamical properties of the local Hubble flow by using high resolution simulations that
include most of the relevant physical processes that lead to the formation of
galaxies inside dark matter haloes.
This paper is organized as follows: in Sect. 2,  we summarize the
numerical methodology; in Sect. 3 we present results on the dispersions of
peculiar velocities around the mean Hubble local flow derived from our
numerical models
and, finally, in Sect. 4, our main conclusions are given.

\section{Numerical modeling}

\subsection{Initial conditions and simulation methodology}

The cosmological simulations analysed in this work use the technique of
``zoom'' (Tormen, Bouchet \& White 1997). We summarize the main steps here.
First, we have performed a cosmological simulation  for  a   $\Lambda$CDM
universe  using WMAP5 parameters (Komatsu et al. 2009), namely
$\Omega_M=0.274$,  $\Omega_{\Lambda}=0.726$, $\Omega_B=0.0456$,  $H_0=70.5$
km/s/Mpc, $n=0.96$ and $\sigma_8=0.812$. 
The simulation was performed in a periodic box of side $100\,h^{-1}$ Mpc
with $512^3$ dark matter particles (e.g. with mass resolution
of $\sim 5.6 \times 10^8\,h^{-1}M_\odot$).
The simulation  started at $z\sim 40$ and ended at the present time $z=0$ where we have
selected one local group (LG) type halo (see section \ref{selection}).
In a second step, we have resimulated this LG candidate by using
an equivalent resolution of $2048^3$ particles within a sphere of $7\,h^{-1}$ Mpc. 
The particle mass
resolution is progressively degraded around this region of interest to reach
the lowest resolution of  $256^3$ effective particles. 
Initial conditions have been generated from the MPgrafic code (Prunet et
al. 2008), a parallel (MPI) version of {\tt Grafic } (Bertschinger 2001).

In order to study effects of baryons, we have run 2 simulations using the
same set of initial conditions: in the first
one, only the dark matter component is considered (referred to as \rundm)
 whereas baryons are included
in the second one (referred to as \runb)  with a full treatment of the physics
of baryons (see below).
In these cases, the highest mass resolution are 
$m_{DM}\approx 8.8\times 10^6\,h^{-1} M_\odot$ (\rundm) and
$m_{DM}\approx 7.4\times 10^6\,h^{-1}
M_\odot$, $m_{gas}=m_{stars}\approx 1.5\times 10^6\,h^{-1} M_\odot$ (\runb).
The Plummer-equivalent force softening adopted for the
high mass resolution particles were 1 and 0.5 $h^{-1}$ kpc
for dark matter and gas particles respectively
 and were kept constant in comoving units.
An intermediate pure dark matter zoom simulation with a lower resolution 
(e.g. $1024^3$ DM effective particles in the central part)
has been realized in order to check
the convergence of the dynamical and physical properties of the cental main
haloes  (masses, relative distance of haloes, etc...).

\begin{figure*}
\rotatebox{0}{\includegraphics[width=16.5cm]{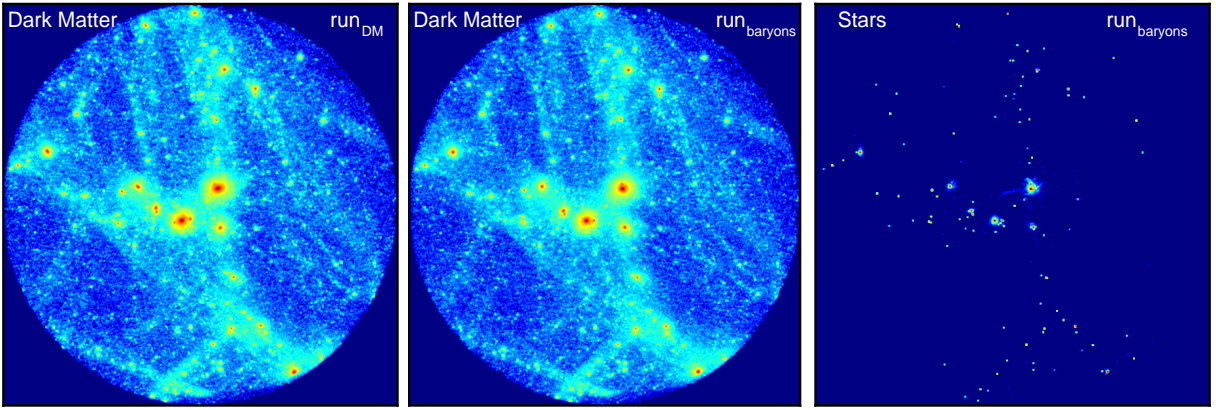}}
\caption{The projected density of dark matter from \rundm \, (left panel) and 
\runb \, (middle panel) within a sphere of radius 4 Mpc. The right panel represents
 the  projected density of stars in the same volume.}
 \label{fig1}
 \end{figure*}

The simulations were performed using GADGET2 
(Springel 2005). As far as the \runb \, is concerned, we have
added  prescriptions for cooling, star
formation, feedback from Type Ia and II supernovae (SN), UV background
and metal enrichment.
Most of this implementation have been tested (effects of mass resolution
on the star formation rate for instance)
and used in previous works (see, for instance, Peirani et al. 2009a,b).
For the sake of clarity, we summarize the main points in the following.

The evolution of the thermodynamic properties of the gas depends on
a  range of physical processes. 
First we use a standard UV-background prescription 
described in Haardt \& Madau (1996).
The cooling and star formation (SF) recipes follow the
prescriptions of Thomas \& Couchman (1992)
and Katz, Weinberg \& Hernquist (1996),
respectively.  Gas particles with $T>10^4$ K
cool at constant density with a given metallicity
 for the duration of a timestep. Gas particles with
$T< 2\times 10^4 K$, number density $n > 0.1\, cm^{-3}$,
overdensity $\Delta \rho_{gas}> 100$, and ${\bf \nabla . \upsilon}
<0$ form stars according to the standard SFR prescription:
$d\rho_*/dt = c_* \rho_{gas}/t_{dyn}$, where $\rho_*$ refers to
the stellar density, $t_{dyn}$ is the dynamical
timescale of the gas, and $c_*$ is the SF efficiency.
Instead of creating new (lighter) star particles, we implement the SF
prescription in a probabilistic fashion.
Assuming a constant,
dynamical time scale across the timestep, 
the probability $p$ that a gas particle
forms stars in a time $\Delta t$ is
$p = 1-\exp(-c_* \Delta
t/t_{\rm dyn})$. For each eligible gas particle, we select a random number
($r$) between 0 and 1 and convert it into a star if $r<p$.

Next, star particles are treated as evolving
stellar populations that inject energy into the
inter-stellar medium, through SN explosions, as
a function of time.
For this, we consider stellar lifetimes in the mass
ranges of $0.8\,M_\odot\,<m<8.0\,M_\odot$ and
$8.0\,M_\odot<m<80.0\,M_\odot$ for Type Ia and Type II progenitors,
respectively. Using a Salpeter initial mass function for Type II
SN, we find that the rate of energy injection is:

\begin{equation}
H_{SN_{II}}=2.5\times10^{-18}\Big(\frac{m_*}{M_\odot}\Big)E_{SN}\Big(\frac{1300}{\tau(\textnormal{Myr})-3}\Big)^{0.24}
\textnormal{erg.s$^{-1}$},
\end{equation}

\noindent
where $E_{SN}=10^{51}$ erg, $m_*$ is the mass of the stellar
population, and $3.53 <\tau <29$ Myr. For Type Ia SN, the heating
is delayed, since they appear $t_0=0.8-1.0$ Gyr after the onset of
star formation. Following de Freitas Pacheco (1998), the
probability of one event on a timescale $\tau$ after the onset of
star formation is given by:

\begin{equation}
H_{SN_{I_a}}=4.8\times10^{-20}\Big(\frac{m_*}{M_\odot}\Big)E_{SN}\Big(\frac{t_0}{\tau}\Big)^{3/2}
\textnormal{erg.s$^{-1}$}.
\end{equation}

Equations (1) and (2) are used to compute the energy released
by SN derived from a star particle $i$ ($E_i$). A fraction $\gamma$ of
this energy is deposited in the j$^{th}$ neighbour gas particle by
applying a radial kick to its velocity with a magnitude $\Delta
v_j = \sqrt{(2w_j\gamma E_i/m_j)}$, where $w_j$ is the weighting
based on the smoothing kernel and $m_j$ is the mass of gas
particle j. We note that all gas neighbours are located in a
sphere of radius $R_{SN}$, centered on the SN progenitor, to avoid
spurious injection of energy outside the SN's region of influence.
In the following, we use the following standard values: $\gamma=0.1$,
$R_{SN}=0.4$ kpc, and $c_* = 0.02$.

\begin{table*}
\begin{center}
\caption{Physical properties of the two central main objects in both runs. Numbers
in parenthesis correspond to number of particles of each component. 
 $D_{MW-M31}$ specifies the physical distance between
the two central objects and the unit of mass is $h^{-1}$ M$_\odot$. } 
\begin{tabular}{ccccccc}
\hline
\multicolumn{1}{c}{} & \multicolumn{1}{c}{\rundm  }& \multicolumn{5}{c}{\runb } \\
\hline
Galaxy & $M_{tot}$ & &$M_{tot}$  & $M_{DM}$ & $M_{star}$&$ M_{gas}$ \\
\hline
MW  & $8.85 \times 10^{11}$&  &$8.96 \times 10^{11}$      &$7.58 \times 10^{11}$ & $1.08 \times 10^{11}$ & $3.02 \times 10^{10}$ \\
    & (100\,097) & & (196\,947)       &(102\,846)           &  (73\,600)           & (20\,501)\\  
M31 & $8.84\times 10^{11}$ & & $8.65 \times 10^{11}$     &$7.43 \times 10^{11}$ & $8.44 \times 10^{10}$ & $3.81 \times 10^{10}$ \\
    & (100\,052) & &(183\,926)      &(100\,856)           &  (57\,223)           & (25\,847) \\
\hline
\multicolumn{1}{c}{D$_{MW-M31}$} & \multicolumn{1}{c}{1.01 Mpc  }& \multicolumn{5}{c}{1.03 Mpc } \\
\hline
\end{tabular}
\label{table1}
\end{center}
\end{table*}

Supernovae do not only inject energy  to the ISM
but also diffuse heavy metals leading to a progressive 
enrichment of their nearby environment. 
To estimate the stellar metallicity, we assume that
the first generation of stars will have ``zero-metallicity'' and
the subsequent generations will be formed in a medium enriched in 
metals by the previous generations. To estimate the metal
abundance, we have considered two elements: Mg (magnesium), representing
the ``$\alpha$-abundance'' and Fe (iron), representing all heavy elements of
the ``iron-peak''. The former is produced essentially by the type II events
whereas the later is produced mainly by type Ia supernovae.

The amounts of Fe and Mg produced by type II supernovae
originating from each stellar cluster are taken from 
Umeda et al. (2002):

\begin{equation}
M_{Fe, Mg} = \big<m_{Fe, Mg}\big>_{SN_{II}} m_* \lambda_{SN_{II}} \,\,\,
\textnormal{$M_\odot$},
\end{equation}

\noindent
where  $m_*$ is the mass of star particle in the simulation.
All Fe (or Mg) is produced after only 29 Myr. Using a Salpeter
IMF, the fraction by mass of these events is $\lambda_{SN_{II}}=5.72 \times 10^{-3} M_\odot^{-1}$
and the average Fe and Mg masses produced by type II supernovae are
$<m_{Fe}>_{SN_{II}}=0.068\,M_\odot$ and $<m_{Mg}>_{SN_{II}}=0.12\,M_\odot$.
In the case of type Ia supernovae, the amount of Fe and Mg ejected after
a given age of a given generation is

\begin{equation}
M_{Fe, Mg} = \big<m_{Fe, Mg}\big>_{SN_{Ia}} m_* \lambda_{SN_{Ia}}\Big[1-\big(\frac{t_0}{\tau}\big)^{1/2}  \Big] \,\,\,
\textnormal{$M_\odot$},
\end{equation}

\noindent
where $\lambda_{SN_{Ia}}=7.2 \times 10^{-4} M_\odot^{-1}$,
$<m_{Fe}>_{SN_{Ia}}=0.85\,M_\odot$ and
$<m_{Mg}>_{SN_{Ia}}=0.026\,M_\odot$ (from model W7 by Nomoto et al. 1997).

The mass of Fe and Mg injected in the medium follows the supernovae
rate estimated previously, being redistributed according to the 
smoothing kernel among the gas particles in the sphere of radius $R_{SN}$.
Then, at each timestep, one can estimate the [Fe/H], [Mg/H] and  [Mg/Fe]
abundances of each gas particle as well as a global metallicity $Z$ by
using the formula 
$logZ=[Fe/H] -1.70 + log(0.362+0.638f_\alpha)$ derived in
Salaris, Chieffi \& Straniero (1993), where
$f_\alpha=10^{[\alpha/Fe]}$ is the enhancement factor
of $\alpha$-elements with respect to iron. 
In our model, the [Mg/Fe] abundance
is  used as a proxy of [$\alpha$/Fe].
The inclusion of metal enrichment processes in the present study 
allows us to improve the treatment of the gas cooling
using metal-dependent cooling functions adapted from
Surtherland \& Dopita (1993).
Note that a similar methodology was adopted in previous works
(see for instance, Kobayashi 2004;
Tornatore et al. 2004; Scannapieco et al. 2005).

\subsection{Properties of simulated  Local Group type objects}
\label{selection}

In this study, we focus on  pairs of galaxies with physical
characteristics similar to the Milky Way-Andromeda pair (MW-M31).
 Each LG group candidate
must satisfy the following criteria:

\begin{enumerate}

\item[(a)] The group contains two main haloes of comparable mass
 ($\sim 0.7-1.2\times 10^{12}\, h^{-1} M_\odot$),
are separated from a distance
 $0.8 \leq D \leq 1.0$ Mpc and their relative radial
velocity is negative.

\smallskip

\item[(b)] There is no object with mass greater than $5 \times 10^{11} h^{-1} M_\odot$ 
and $10^{13} h^{-1} M_\odot$
within the distance $3$ and $8$ Mpc from the center of mass of the main pair of haloes
respectively. 

\smallskip
\item[(c)] There is only one Virgo-like halo ($M\geq 10^{14}\, h^{-1} M_\odot$)
 at the distance of $8$ to $13$ Mpc.
\end{enumerate}

Such selection criteria are quite similar to those used in
previous works (Macci{\`o}, Governato \& Horellou 2005; Hoffman
et al. 2008; Tikhonov \& Klypin 2009; Martinez-Vaquero et al. 2007, 2009)
and then make the comparison of the results easier. 
Several candidates of pairs of dark matter haloes
that satisfy criterion (a) can be easily found but almost all of them
are ruled out by criteria (b) and (c).
In the end,  only one candidate that satisfy all of the above criteria has been
found. This latter system has been resimulated with and without baryons 
and using higher resolutions.
The simulations were performed on local parallel supercomputers
at IAP (France) and the CPU times
for \rundm \, and \runb \, were $\sim$ 2 days and $\sim$ 2.5 months respectively
using 32 processors. 
Table \ref{table1}
summarizes the main properties of the central objects in
both runs and the number of particles considered for each component. 
In the following, we will mostly characterize dark
matter haloes by their maximum circular velocity which can defined by 
$V_{max}=max(\sqrt{Gm(\leq r)/r}\,)$, where $m(\leq r)$ is the mass
enclosed at radius r. $V_{max}$ can be estimated without any accurate 
estimate of the physical boundary of the objects which can be proved
to be difficult especially for subhaloes.

Fig. \ref{fig1} shows the projected density of dark matter from \rundm \,
and \runb \,
as well as the projected density of stars within a sphere of physical
radius 4 Mpc.
The distributions of dark matter look very similar between the two simulations.
However, the dark matter haloes seem to be more concentrated
when baryons are included in the simulation. This can be explained by the 
fact that dissipation of the gas (from radiative cooling processes) and 
subsequent star formation lead to a steeper DM density profile due
to adiabatic contraction.
This process can also be observed  in dark matter
substructures which become less sensitive to tidal torques and will
survive a longer time in their host haloes, as has been extensively
discussed in the past literature (Blumenthal et al. 1986; Gnedin et al. 2004;
Duffy et al. 2010 and references therein).

\subsection{Mock catalogs of dark matter haloes and galaxies}
\label{comparison}

One of the main purposes of this work is to study  how the inclusion
of baryons may affect the
dynamical properties of the simulated local universe. Then,
we have used a simple algorithm which allows  a direct comparison between
the two simulations.

First, we have prepared mock catalogs of dark matter haloes and galaxies at $z=0$
for each simulation. To do this we have used the public code
AMIGA\footnote{http://popia.ft.uam.es/AMIGA} (Knollmann \& Knebe 2009) 
which is an efficient halo and substructure finder
by identifying  local density maxima in an adaptively smoothed density field.
The use of this code is also attractive since
subhaloes are separated form their parent haloes and
 each (sub)halo center is estimated accurately by  
an iterative algorithm, which is crucial 
in this study.
For each object, we define the virial radius $R_V$ (and thus the virial
mass $M_V$) as the radius where
the enclosed mean density $M_V/(4\pi R_V^3/3)$
is $\Delta_c$ times the critical density,
$\rho_c(z)=3H(z)^2/8\pi G$, where $H(z)=H_0\sqrt{\Omega_m(1+z)^3+\Omega_\Lambda}$.
$\Delta_c$ can be computed from the spherical top-hat collapse model, and
in the case of a flat cosmology with $\Omega_m + \Omega_\Lambda =1$, one 
can use the fitting formula from Bryan \& Norman (1998) which
depends on both cosmology and redshift:

\begin{equation}
\Delta_c = 18\pi^2 + 82x-39x^2,
\end{equation}

\begin{equation}
x=\Omega_m(z)-1,
\end{equation}

\begin{equation}
\Omega_m(z)=\Omega_m(1+z)^3\Big(\frac{H_0}{H(z)}\Big).
\end{equation}

\noindent
In the cosmology used for the present study, $\Delta_c=97.6$ at $z=0$.
Then, only bound structures with at least 100 dark matter particles  
(e.g. a mass higher than $8.8\times 10^8\,h^{-1}M_\odot$ in \rundm)
 were retained in the different catalogs.
In the following, {\it SPH haloes}  refers to  objects identified in  \runb \,
which consists of a single dark matter halo containing  a fraction of gas and (or not)
a  {\it galaxy},  namely a bound structure with at least 80 star particles 
(corresponding to a mass of $1.2\times 10^8\,h^{-1}M_\odot$).

\begin{figure}
\rotatebox{0}{\includegraphics[width=\columnwidth]{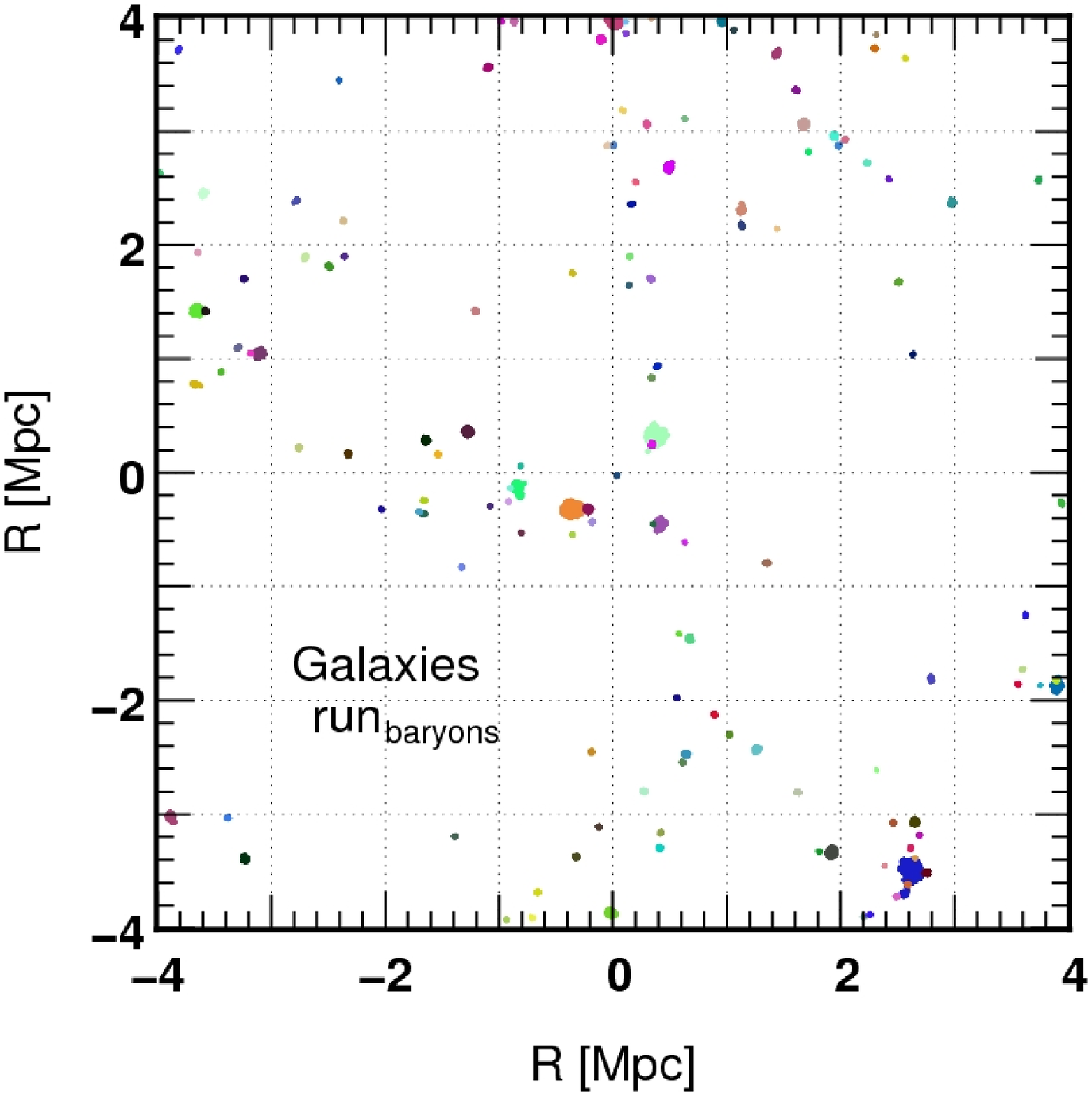}}
\rotatebox{0}{\includegraphics[width=\columnwidth]{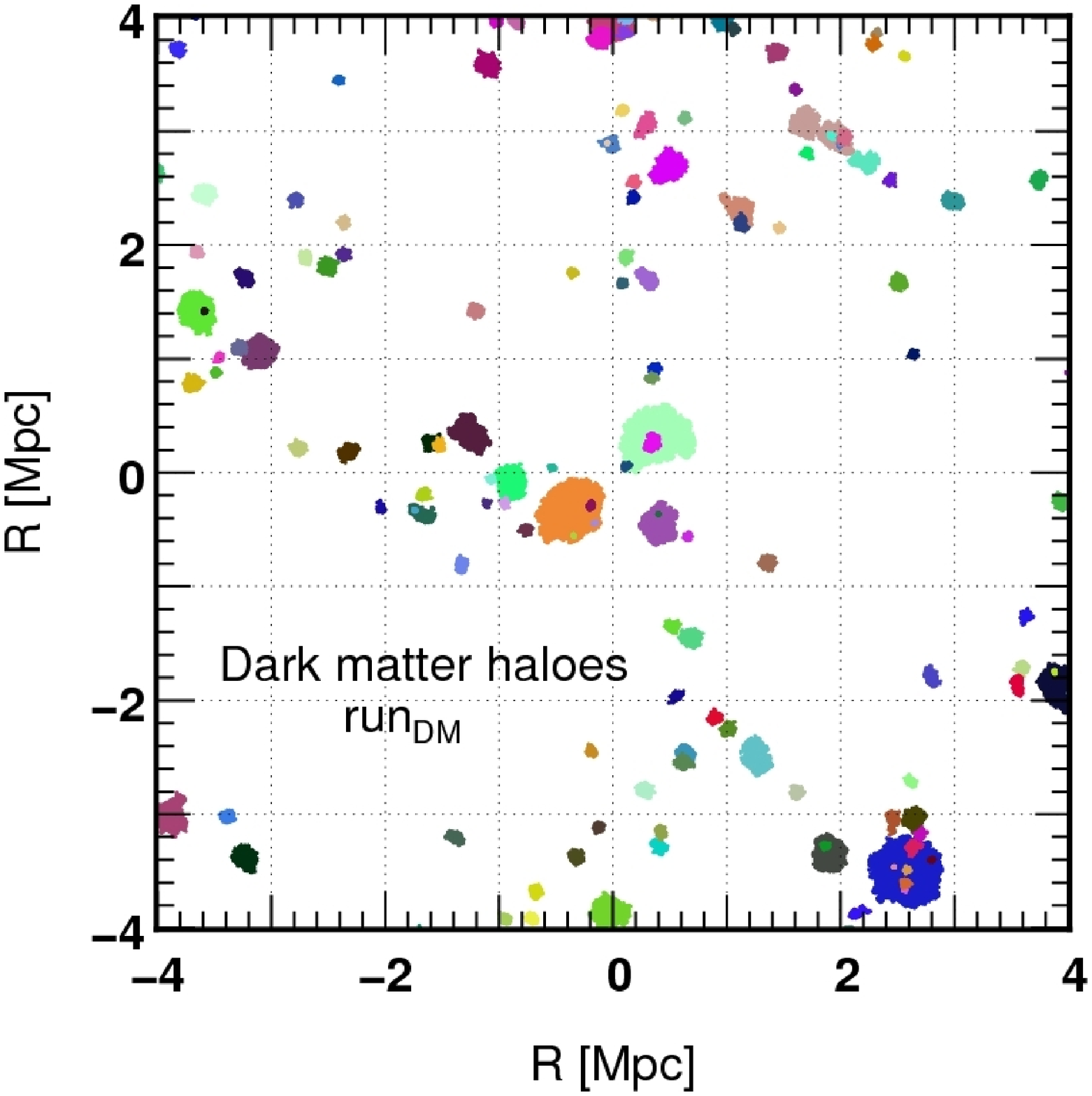}}
\caption{Distributions of galaxies (upper panel) and their corresponding dark
matter haloes in \rundm (lower panel)  and using the same color code. For clarity reason
we have limited to a distance of 4 Mpc from the LG mass center.}
 \label{fig_dist}
 \end{figure}

Once the two catalogs are built, we match the SPH haloes to their equivalent
DM haloes of \rundm \, by using 
the following algorithm.  
Since each particle in the simulation can be
identified, it is
possible to obtain the constitution of each SPH and DM halo at $z=0$. If more than
50\% of dark matter particles of a given SPH halo is found in
a dark matter halo of \rundm,
we assume that both objects are the same.
To illustrate this, Fig. \ref{fig_dist} shows the distribution of all galaxies
that have a corresponding dark matter halo in \rundm \, within a distance of 4 Mpc
from the LG mass center. 
However, it is worth mentioning that 
a small number of SPH haloes have no corresponding dark matter halo in \rundm.
These SPH haloes are mainly substructures belonging to the most massive objects
such as the MW or M31 haloes. In \rundm, they don't
exist anymore, mainly because they have been completely disrupted by tidal torques. 
Such a statement has already been studied in great detail in previous works
(see for instance, Nagai \& Kravtsov 2005; Libeskind et al. 2010,
 Klimentowski et al. 2010).

\begin{figure}
\rotatebox{0}{\includegraphics[width=8cm]{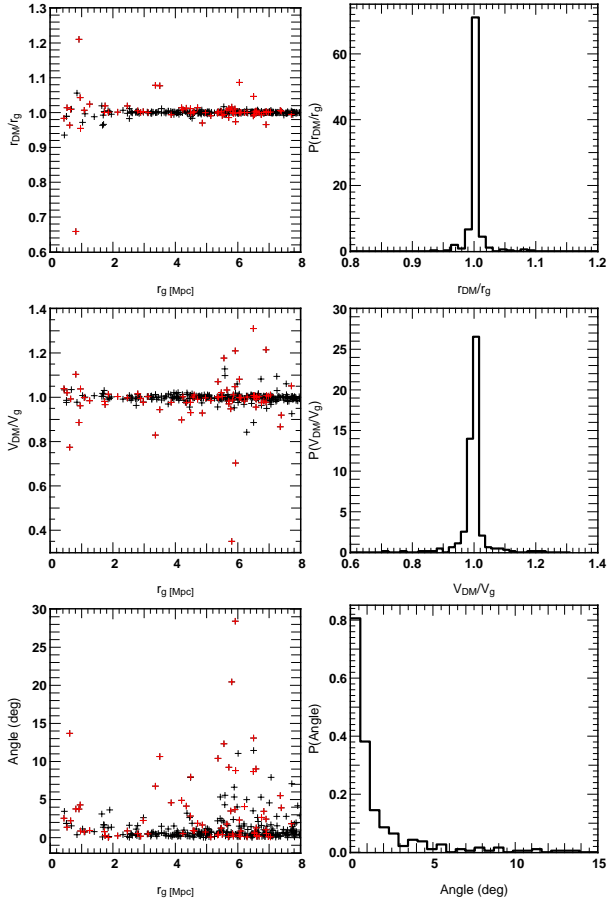}}
\caption{Comparison between  the radial distances  $r_g$ (in the frame
of the Local group)
 and the peculiar velocities $V_g$ of galaxies  
  identified in \runb \, and  those of their corresponding
dark matter haloes  in  \rundm \, ($r_{DM}$,$V_{DM}$)  at $z=0$: 
the first column shows the variations of the norm ratio $r_{DM}/r_g$ (first line),
 the velocity norm ratio $V_{DM}/V_g$ (second line)
 and  angle between velocity vectors (third line) as respect to $r_g$ 
while the probability distribution functions of  $r_{DM}/r_g$ (first line),
 $V_{DM}/V_g$ (second line) and the angle (third line) are plotted in the second column.
The red crosses correspond to dark matter subhaloes with a mass lower that $5\times 10^{9}M_\odot$.
}
\label{fig_prob}
 \end{figure}

It is now interesting to study the statistical differences between the radial distances 
(in the frame of the Local Group)
 and peculiar velocity of galaxies
 ($r_g$,$\overrightarrow{V_g}$) identified from \runb \, and  those of their corresponding
dark matter haloes  in  \rundm \, ($r_{DM}$,$\overrightarrow{V_{DM}}$) at $z=0$.
The main results are shown in Fig. \ref{fig_prob}  where
the probability distribution functions of the position ratio $r_{DM}/r_g$, the
velocity norm ratio $V_{DM}/V_g$ and the angle
($\overrightarrow{V_g}$,$\overrightarrow{V_{DM}}$)  are plotted.
Fig. \ref{fig_prob} also shows the
  variations of $r_{DM}/r_g$,  $V_{DM}/V_g$ and
 ($\overrightarrow{V_g}$,$\overrightarrow{V_{DM}}$) with respect to positions $r_g$.
 It appears clearly that most
of the galaxies have very similar position and velocity properties to their corresponding
dark matter haloes. However, in some cases, we  notice important differences between 
the two runs. This happens mainly for substructures  ($\sim5\times 10^{9}M_\odot$) in
more massive objects where their dynamics and orbits are more likely to be modified by
the halo potential, as previously observed by Weinberg et al. (2008).
Moreover, a detailed study by Libeskind et al. (2010)
has also shown  that subhaloes in hydrodynamic runs tend to be more radially
concentrated than those in pure dark matter ones.

\begin{figure}
\rotatebox{0}{\includegraphics[width=\columnwidth]{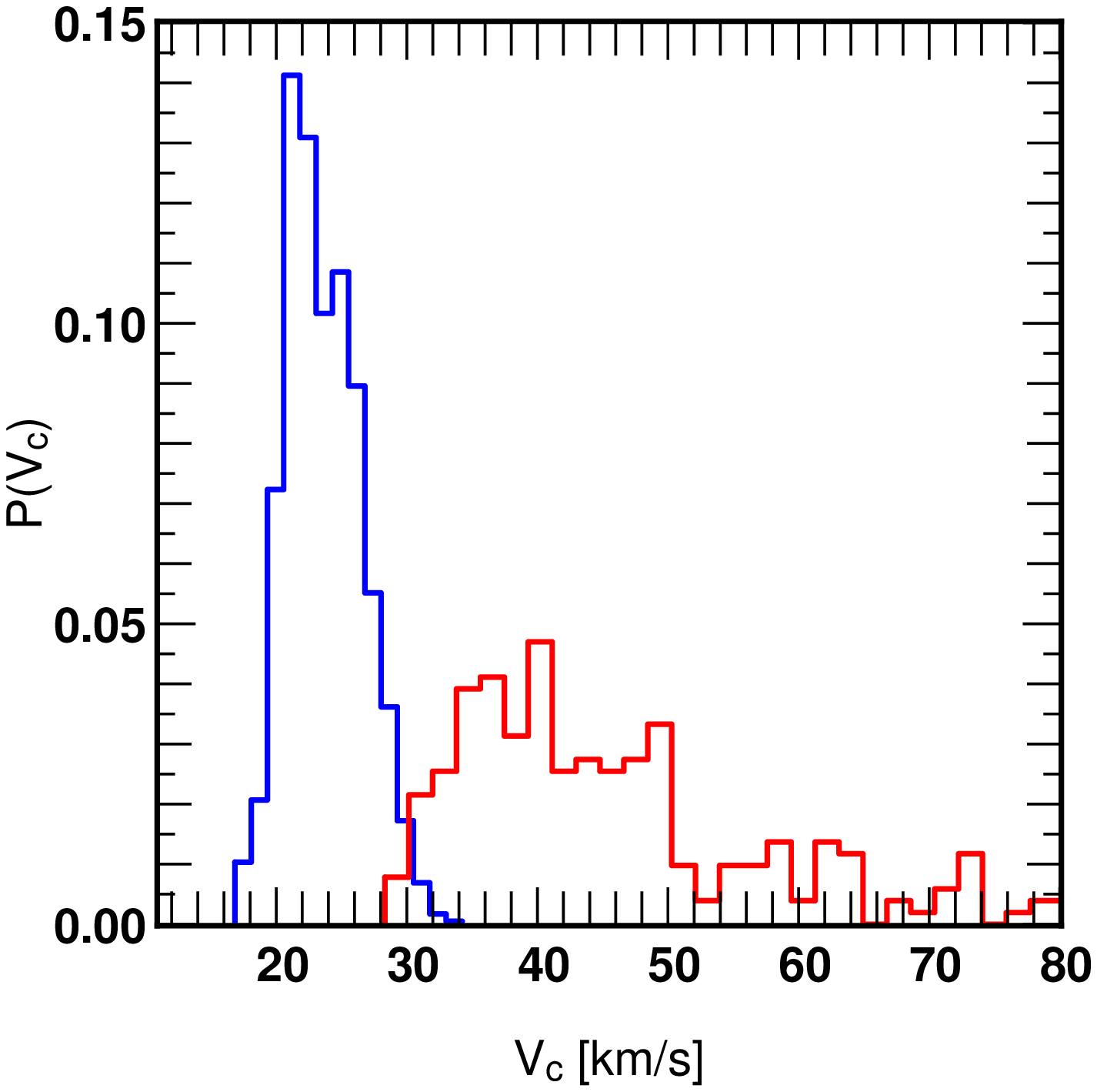}}
\rotatebox{0}{\includegraphics[width=\columnwidth]{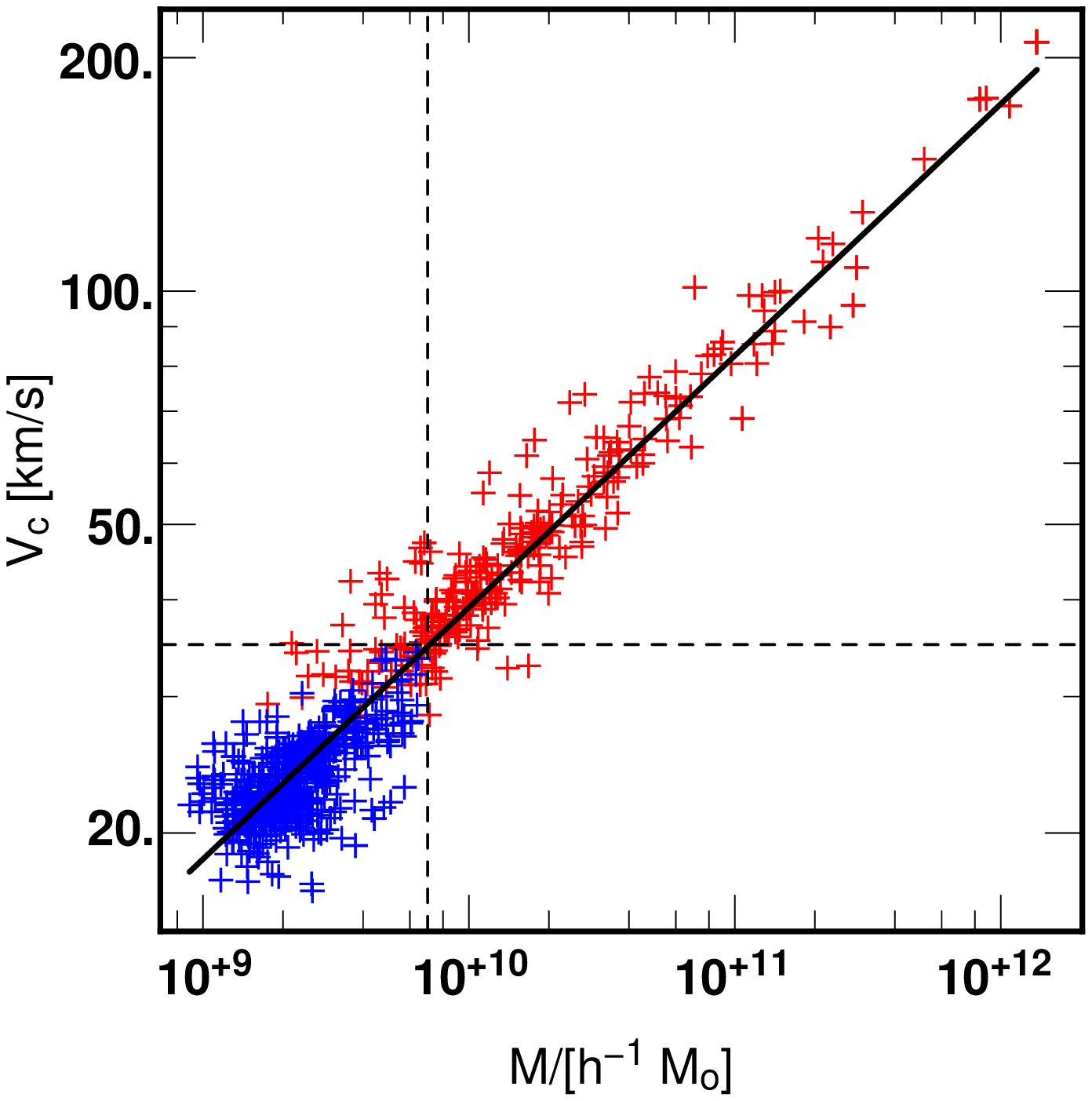}}
\caption{Dark matter haloes from \rundm: the probability distribution
 functions of their maximum circular
 velocity $V_c$ (upper panel) and the variations of their mass as respect
 to $V_c$ (lower panel). Blue color refers to
 DM haloes that have corresponding starless DM haloes in \runb \, whereas
 red color refers to DM  haloes that match SPH haloes with a galaxy. In the
lower panel, the solid line represents our best fit (see the text for more details)
while the dashed lines show $V_c=35$ km/s and  $M=7\times 10^{9}\,h^{-1}M_\odot$.}
 \label{fig_vc_dm}
 \end{figure}

Among the population of SPH haloes, there is a
subpopulation of dark matter haloes that contain no stars and thus
no galaxy. From a theoretical point of view, the existence of such
dark halo populations can be explained  
 by the fact that their star
formation have been suppressed  by feedback processes such as
UV-background (Thoul \& Weinberg 1996; 
Bullock, Kravtsov \& Weinberg 2000; Somerville 2002;
Ricotti, Gnedin \& Shull 2002; Benson et al. 2002; Read, Pontzen \& Viel
2006; Hoeft et al. 2006; Okamoto, Gao \& Theuns 2008). 
In particular,  the presence of a
photoionizing background suppresses the formation of galaxies with circular velocities
$V_{circ}\leq 30$ km/s  (Thoul \& Weinberg 1996; Bullock, Kravtsov \& Weinberg 2000)
and a similar result has been  derived by Tikhonov \& Klypin
(2009) in order to explain the observed size of local voids,
using models with $\sigma_8=0.75$. Moreover, 
Hoeft et al. (2006) have also studied  the formation of dwarf galaxies in voids 
using high-resolution hydrodynamical simulations including UV-background.
They derived a critical mass for haloes at $z=0$, $M_c=6.5\times 10^{9}\,h^{-1}M_\odot$,
below which the  cooling of gas is suppressed and  similar results have
been derived by Okamoto, Gao \& Theuns 2008).
Then, we have compare  the properties of dark matter haloes of \rundm \,
that have  either a corresponding galaxies or a dark halo in \runb.
In Fig. \ref{fig_vc_dm}, the probability distribution function is represented for
the maximum circular velocity $V_c$ of DM haloes from \rundm \, and the
variations of $V_c$ as respect to
their total halo mass. The results suggest that
there is a critical value ($V_c\sim 30$ km/s) where the star formation is
suppressed, in good agreement with past works. Moreover,  dark matter haloes that
have corresponding starless SPH haloes in
\runb \, are characterized by $V_c\leq 35$ km/s and have a mass lower than
$\sim 7\times 10^{9}\,h^{-1}M_\odot$. 
Note that, the circular velocity of dark matter haloes of \rundm \, can be well approximated
by   $V_c=39.08\, $km$\,s^{-1} \times (M/10^{10}h^{-1}M_\odot)^{0.32}$ in
good agreement with Hoeft et al. (2006) and Klypin, Trujillo-Gomez \& Primack (2010).

\begin{table*}
\begin{center}
\caption{The velocity dispersion around the Hubble flow  $v_H(r)=H_0r$ for different samples of
  selected galaxies (\runb) or dark matter haloes  (\rundm) as a function of
  the distance from the mass center of the simulated MW-M31 pairs. Each number in
parenthesis gives the number of objects that have been considered in the 
computation of $\sigma_H$.} 

\begin{tabular}{cccccccc}
\hline
R(Mpc) &   $\sigma_*$$^a$ & $\sigma_{*,cor}$$^b$ & $\sigma_{*\leftrightarrow DM}$$^c$ & $\sigma_{DM\leftrightarrow *}$$^d$  &  $\sigma_{DM, V_c \geq 45}$$^e$  &  $\sigma_{DM, V_c \geq 30}$$^f$  &  $\sigma_{DM, V_c \geq 20}$$^g$ \\
\hline
1-2 & 51.4 (17)  & 50.1 (17) & 51.4 (17) & 52.8 (17) & 48.5 (7) & 52.0 (20) &59.1 (59) \\
1-3 & 45.2 (37)  &  42.4 (37)& 45.1 (36) & 44.3 (36) & 39.0 (14)& 44.1 (41) &47.3 (126) \\
1-4 & 58.5 (66)  & 54.8 (66) & 57.5 (64) & 56.2 (64) & 48.2 (21)& 53.9 (72) &57.0 (217) \\
1-5 & 66.9 (125) & 61.4 (125)& 61.9 (114)& 61.6 (114)& 69.0 (48)& 64.8 (134)&65.0 (372) \\
1-6 & 75.1 (188) & 68.2 (188)& 72.3 (172)& 71.1 (172)& 74.3 (69)& 72.8 (192)&70.1 (532) \\
1-7 & 83.5 (267) & 75.2 (267)& 81.4 (246)& 80.7 (246)& 82.7 (92)& 79.9 (269)&77.6 (724) \\
1-8 & 98.9 (331) & 90.4 (331)& 98.5 (308)& 97.9 (308)& 101.8 (120)&94.7 (332)&91.5 (904) \\
\hline
$rms^h$ & 0.0 & &2.3 &3.0 &4.9 &3.0 &5.1\\
\hline
\end{tabular}
\label{table3}
\end{center}
\vspace{-0.25cm}
{\small
\hspace{-6.0cm}$^a$$\sigma_H$ computed from galaxies
identified in \runb \, (Sample $A$).

\hspace{-1.45cm}$^b$$\sigma_H$ computed from galaxies
identified in \runb \, (Sample $A$) but corrected from distance errors.

\hspace{0.3cm}$^c$$\sigma_H$ computed from galaxies
identified in \runb \, that have a corresponding DM halo in  \rundm  \, (Sample $B_1$).

\hspace{0.42cm}$^d$$\sigma_H$ computed from DM haloes 
identified in \rundm  \, that have a corresponding galaxy in  \runb \, (Sample $B_2$).

\hspace{0.4cm}$^e$$\sigma_H$ computed from DM haloes 
identified in \rundm  \, with a maximum circular velocity $V_c\geq$45 km/s (Sample $C_1$).

\hspace{0.4cm}$^f$$\sigma_H$ computed from DM haloes 
identified in \rundm  \, with a maximum circular velocity $V_c\geq$35 km/s (Sample $C_2$).

\hspace{0.4cm}$^g$$\sigma_H$ computed from DM haloes 
identified in \rundm  \, with a maximum circular velocity $V_c\geq$20 km/s (Sample $C_3$).

\hspace{-10cm}$^hrms=\sqrt{\frac{1}{7}\sum\big(\sigma_H^2(R) - \sigma_{*}^2(R)\big)}$
}

\end{table*}

\section{The local Hubble flow}

The local Hubble flow  has proved to be cold
according to past observational analysis.
 As mentioned in the
introduction, this puzzle has been studied in various numerical works but no one
so far has investigated the impact of baryons, the main purpose of this work. 

In the following,  the peculiar velocity dispersion $\sigma_H$
in the local volume is estimated by using the root mean square (RMS) of the residual from
the global Hubble flow, defined by:

\begin{equation}
\sigma_H^2=\frac{1}{N}\sum_{i=1}^{N}\big(v_i-v_H(r_i)\big)^2
\end{equation}
\noindent

\noindent
where  $v_i$ and $r_i$ are  radial velocities and radial distances of galaxies (or dark
matter haloes) in the frame of the Local Group where the origin matches with
the mass center of the MW-M31 pair, and $v_H$ is the global Hubble flow. 
In order to compare with recent works, and more specifically with the analysis
of  Tikhonov  \& Klypin (2009) and Martinez-Vaquero et al. (2009),
we have estimated $\sigma_H$ using different samples of dark matter haloes 
or galaxies  with distances between  1 to D Mpc
(where D$=$2, 3, 4, 5, 6, 7 and 8 Mpc) and considering $v_H(r_i)=H_0r_i$,
with $H_0=70.5$ km/s/Mpc. However, it is worth mentioning that
 for distances within  $\sim$ 3 Mpc,
 due to the decelerating influence of the main
central objects to nearby low mass objects, the Hubble flow is not well
described by  $v_H(r_i)=H_0r_i$  (see Fig. \ref{fig_h_star}) 
and one should use  another relation such as
the one derived in Peirani \& de Freitas Pacheco (2008).
Then,  the study has been limited to a
distance of 8 Mpc to avoid to be in the vicinity of the boundary 
of the highest resolution domain.
First, sample $A$  includes  all galaxies identified in \runb \,
and represents therefore the fiducial sample. For this case only,  $\sigma_H$
has been estimated either by assuming a 10\% error as the mean
error of distance measurements (referred to as $\sigma_{*,cor}$) or with no error
(referred to as $\sigma_{*}$). 
Samples $B_1$ and  $B_2$ regroup galaxies
and their  corresponding dark matter haloes in \rundm \, respectively and have
therefore the same number of objects which permits a direct comparison between
the two runs. 
Finally, samples $C_1$, $C_2$ and $C_3$ gather
dark matter haloes  extracted from \rundm \, 
with circular velocities higher than 45, 30 and 20 km/s respectively. 

\begin{figure}
\rotatebox{0}{\includegraphics[width=\columnwidth]{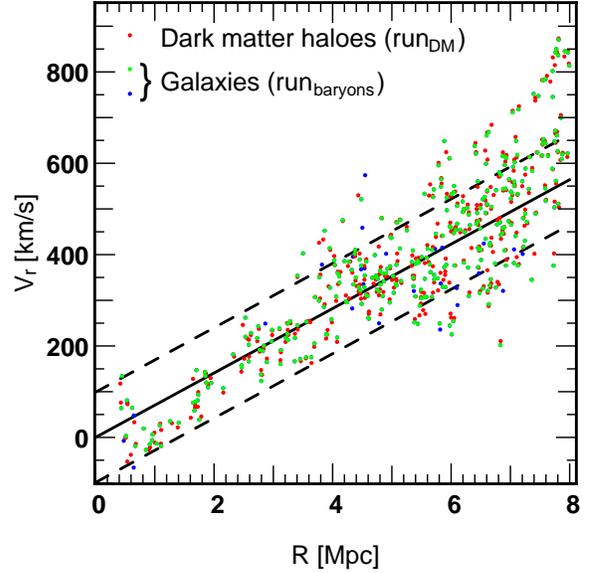}}
\caption{Simulated radial velocity-distance data for galaxies of sample $B_1$ (green dots) and
 dark matter haloes from sample $B_2$ (red dots). The blue dots correspond to
 velocity-distance data for galaxies that don't have a corresponding halo in
 \rundm. The solid line represents the $H_0r$ Hubble flow and the dashed ones
 show $H_0r \pm 98.9$ km/s.}
 \label{fig_hubble}
 \end{figure}

\begin{figure}
\rotatebox{0}{\includegraphics[width=\columnwidth]{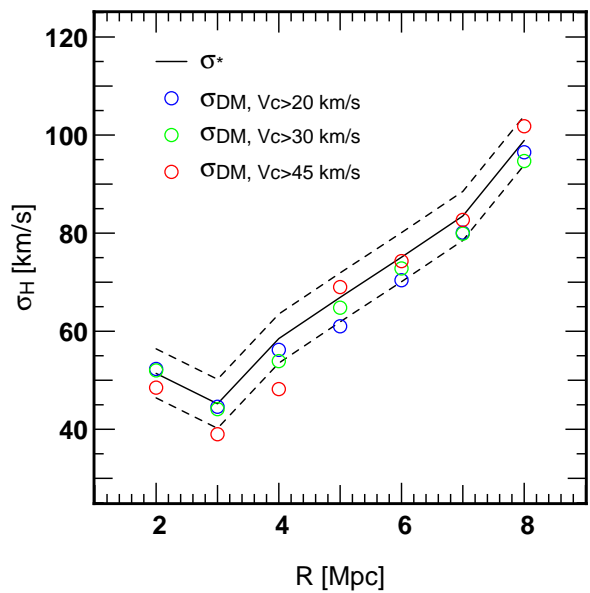}}
\caption{The velocity dispersion around the Hubble flow $v_H(r)=H_0r$ as a function of
  the distance from the mass center of  simulated MW-M31 pairs
for galaxies identified in \runb \, ($\sigma_*$, black line), 
 and for
dark matter haloes from \rundm \,with a maximum circular velocity
defined by $V_c \geq 20$ km/s (blue circles), 
$V_c \geq 30$ km/s (green circles),
$V_c \geq 45$ km/s (red circles). The dashed lines show
$\sigma_* \pm 5$ km/s.}
 \label{fig_sigma}
 \end{figure}

\begin{figure}
\rotatebox{0}{\includegraphics[width=\columnwidth]{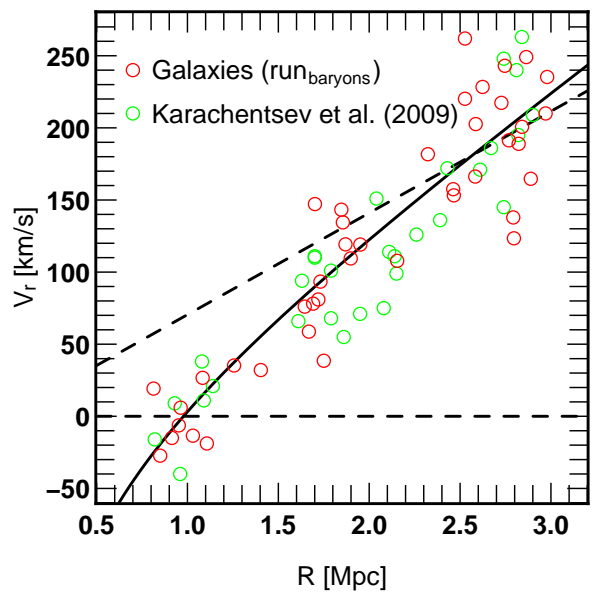}}
\rotatebox{0}{\includegraphics[width=\columnwidth]{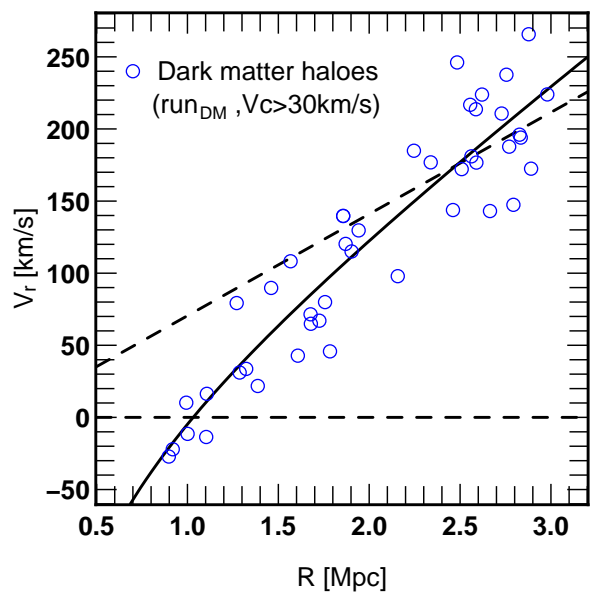}}
\caption{Upper panel: radial velocities and distances for observed (green circles) and simulated
(red circles) galaxies in the neighborhood of the LG with respect to its centroid.
In the lower panel, 
the velocity and distance data derived from dark matter haloes in \rundm \, with $V_c\geq 30$ km/s
is shown. In both panels, the inclined dashed line represents
the linear Hubble relation with $H_0=70.5$ km/s while the solid line represents
the best fit to the $v_r=v_r(r)$ relation derived in Peirani \& de Freitas Pacheco (2008).
}
 \label{fig_h_star}
 \end{figure}

Fig. \ref{fig_hubble} shows the Hubble diagram for galaxies and dark matter haloes
belonging to sample $A$, $B_1$ and $B_2$. As expected from results presented in the previous
section,  radial distances and velocities 
of galaxies and those of their corresponding dark matter haloes in \rundm \,
follow the same trend. Such a statement is also confirmed by
Table \ref{table3} where all the  $\sigma_H$ values
derived at different distances from the LG mass center
and for different samples are shown.
In particular, $\sigma_H$ values derived from samples  $B_1$ and $B_2$ are very close
and suggests that the presence of  baryons does not affect the velocities dispersion
of satellites around the main pair at distances  1-8 Mpc.
Moreover, results from  samples $C_1$, $C_2$ and $C_3$ show  that 
estimations of  $\sigma_H$ from the pure dark matter simulation
give similar values than those derived from the fiducial sample as it can
been clearly seen in Fig. \ref{fig_sigma}.
For instance, when dark haloes are selected with $V_c\geq 30$ km/s, the number of
objects considered is quite close to the ``true'' number of galaxies and the mean error
on $\sigma_H$ is in order of  $\sim 3$ km/s.
However, when haloes are selected with $V_c\geq 45$ km/s and
 $V_c\geq 20$ km/s, both cases lead to an estimation of  
$\sigma_H$ with a mean error of  $\sim 5$ km/s as respect to $\sigma_*$ values although 
the number of objects taken into account for the $\sigma_H$
computation are  underestimated or overestimated
respectively.

To finish,  it is instructive to compare  the velocity dispersion values derived from
our numerical model
to $\sigma_H$ values obtained from the analysis by Tikhonov  \& Klypin (2009)
who have considered the  most complete observational
data using an updated version of the
Karachentsev et al. (2004) sample of galaxies. The number of galaxies used to
compute $\sigma_*$ (first column of Table \ref{table3}) is in quite good agreement
with number of galaxies considered  within 1-8 Mpc (see their Table. 3). However, we found
that  $\sigma_H$ values derived in this work tend to be
($\sim10$\%) smaller within 1-8 Mpc.
 We have also compared in Fig. \ref{fig_h_star} our simulated data to
updated data on distances and velocities of galaxies 
for distances $0.7\leq D \leq 3$ Mpc without an additional dipole component
(Karachentsev et al. 2009). The nice agreement between simulated data and observations lead to
similar values of the dispersion of peculiar velocities.
 Indeed, using in this case $v_H(r)=V_r(r)$ derived in Peirani \& de Freitas Pacheco (2008),
we found $\sigma_H=29.3$ km/s from observational data whereas
$\sigma_*=33.0$ km/s and $\sigma_{DM,V_c\geq30}=29.9$ km/s  respectively for galaxies
from \runb \, and dark matter haloes from \rundm \, with $V_c\geq 30$ km/s.
But an important feature of  the Hubble flow at such distances
is the decelerating influence of the central main objects on nearby galaxies which lead
the radial velocities-distances regression to cross the line of zero velocity at
the  turnaround radius R$_0$. In this regard,
Lynden-Bell (1981) and Sandage (1986) proposed an alternative method to the virial relation in order
to estimate the mass of the Local Group, which can be extended to other systems dominated either by
one or a pair of galaxies. Based on 
the Lema\^itre-Tolman (LT) model (Lema\^itre 1933; Tolman 1934),
they show that, if the velocity field close to 
the main central body (probed by satellites) allows the determination of R$_0$, then the total
central mass can be 
calculated straightforwardly from the relation

\begin{equation}
M = \frac{\pi^2 R_0^3}{8GT_0^2},
\label{equ_TL}
\end{equation}
where T$_0$ is the age of the universe and G is the gravitational constant. 
In a flat cosmological model with a cosmological constant $\Lambda$,
equation (\ref{equ_TL}) needs to be modified as follows (Karachentsev et al. 2007):

\begin{equation}
M = \frac{\pi^2 R_0^3 H_0^2}{8GT_0^2 f(\Omega_\Lambda)},
\label{equ_TL2}
\end{equation}

\noindent
where

\begin{equation}
f(\Omega_\Lambda) = \frac{1}{\Omega_\Lambda} -\frac{1-\Omega_\Lambda}{2\Omega_\Lambda^{3/2}}cosh^{-1}\Big(\frac{2}{1-\Omega_\Lambda}-1\Big)
\label{equ_TL3}
\end{equation}

We have estimated $R_0$ from our two mock catalogs of galaxies and haloes
by fitting the  $V_r=V_r(r)$ formula derived in Peirani \& de Freitas Pacheco (2008)
to our simulated data. 
From the best fit solutions, we obtained  $R_0= 0.98$ Mpc (galaxies sample) and
$R_0= 1.03$ Mpc (for dark matter haloes sample with  $V_c\geq 30$ km/s).
Using $\Omega_M=0.274$ and  $H_0=70.5$ km/s/Mpc, one can derive  from Eq. (\ref{equ_TL2})
the total masses enclosed at  radius $R_0$, namely
$M=2.0\times 10^{12} M_\odot$ (galaxy sample)
and  $M=2.3\times 10^{12} M_\odot$ (dark matter haloes sample) respectively. 
However, the total mass enclosed in a radius of 1 Mpc is $\sim 3.4 \times 10^{12} M_\odot$ from
both simulations and suggest that masses derived from Eq. (\ref{equ_TL2})
are underestimated by a factor 1.5-1.7.

\section{Discussion and Conclusions}

In the present paper, we have used cosmological N-body simulations with and
without a complete treatment of the physics of baryons  to study  the
formation of a local group type halo. 
We have first identified a LG system in a simulation with a low mass particle resolution,
using standard criteria. Then, the best candidate has been re-simulated with
higher resolution using the technique of zoom.
 This approach differs from the technique of constrained simulations used for instance
in the CLUES project (Yepes et
al. 2009), which include the correct motion
and position of objects in a large volume such as the Local Supercluster, the
Virgo and the Coma cluster and the Great Attractor (see also Lavaux 2010).

We focus on the dynamical properties of the local Hubble flow and in particular, the
influence of  baryons.
From the two runs, we  built mock catalogs of either galaxies or dark matter haloes and computed
their velocities as respect to the mass center of the central main pair (Milky Way-M31).
We found that the dispersions of
peculiar velocities around the mean
local Hubble flow of galaxies in \runb \, and those of their corresponding dark matter
haloes  in \rundm \, are very close for distances $D=1$ to 8 Mpc.
Moreover, similar $\sigma_H$ values (with mean errors $\leq 5$ km/s)
 are also obtained for samples of dark matter haloes
extracted from \rundm \, with constraint of their maximum circular velocity. 
These results  suggest that the global dynamical properties of the Hubble flow
is not affected by the presence of
baryons. Such a result is not very surprising since galaxies are expected to form
in the gravitational potential well of dark matter haloes. Therefore, 
their  properties are expected to be closely related to those of
their host haloes. This statement is confirmed on large scales  as
 shown in Figs. \ref{fig_dist} and \ref{fig_prob}.
However, some discrepancies between positions and dynamical properties of galaxies and
those of their corresponding haloes may raised in the case of
low mass objects in denser environment.
For instance,   the presence of  baryons  is supposed to favor the survival of substructures
inside massive haloes  against tidal striping
 (see Duffy et al. 2010 and references therein) and can affect their
distribution (Weinberg et al. 2008; Libeskind et al. 2010, Knebe et al. 2010). 
But these effects seem to have no 
particular consequences in the estimation of the global $\sigma_H$ values.
It should also be pointed out that \runb \ does
not include any prescriptions of strong feedback
such as galactic winds or Sedov shock waves that
are expected to lead to a significant ejection of
gas from the central region of forming galaxies.
As a result, the fraction of baryons (in the form
of stars and cold gas) in the central region of
simulated massive galaxies such as the Milky Way
or M31 is close to the cosmological baryonic fraction
(see Table \ref{table1}) whereas observations suggest lowest values 
(Fukugita, Hogan \& Peebles 1998;
Bell et al. 2003). However, no one has solved this
“missing baryon problem” yet and recent studies
even suggest that theoretical sources of strong
winds such as AGN and supernovae are insufficient
 to explain the missing baryons in Milky
Way type galaxies (see for instance Anderson \&
Bregman 2010; Silk \& Nusser 2010). Although
the present study mainly focus on lower mass
galaxies ($\leq 10^{10} M_\odot$) in which supernovae feedback
 is supposed to have a significant impact in
their formation, the other sources of strong feedback should be taken into account as well. Then,
a natural extension of this work would be to add
these processes while increasing the mass resolution in the simulation (see
for instance
Ceverino \& Klypin 2009) to see whether the conclusions presented here
(especially the distribution of galaxies as respect to their corresponding dark
matter haloes) are significantly affected of not.

In good agreement with previous theoretical and numerical works, we found
 that the halo population in the pure dark matter run can be
divided into two groups according to their maximum circular velocity $V_c$. There is
indeed a critical value, $V_c\sim 30-35$ km/s (corresponding to
haloes with mass lower than $\sim 7\times 10^{9}\,h^{-1}M_\odot$ in \rundm), below which the
corresponding SPH haloes don't host any galaxy.
Such a phenomenon may be explained by the fact that
the cooling of gas is expected to  be suppressed 
by feedback processes such as a UV-photoionisation in low mass haloes
 (see references in paragraph 2.3).
As regards this point, we cannot exclude the fact that the numerical resolution  
used in this work may affect the results especially for low mass systems.
However,  haloes with mass of $\sim 7\times 10^{9}\,h^{-1}M_\odot$ are composed
by $\sim 800$ particles and this seems reasonable for the  estimation of
the thermodynamic properties of the gas component.  
Then, table \ref{table3} indicates that the number of
dark matter haloes in \rundm \ selected with $V_c\geq 30$ km/s is actually
quite close to the real number of galaxies selected \runb. Therefore, one can use this 
single criteria to select dark matter haloes in a collisionless simulation 
to have an good estimation of $\sigma_H$.
The existence of a dark halo population in the Local Group can also have 
important consequences in understanding the discrepancies 
between the large number of  subhaloes present in simulations
but not observed  (Kauffmann, White \& Guiderdoni 1993;
Moore et al. 1999; Klypin et al. 1999). This problem
is beyond the scope of this paper and will be investigated
in detail in a forthcoming paper.

The velocity dispersion obtained from our fiducial model (e.g. from galaxies)
is in quite good agreement with observational expectations.
For instance, with added corrections for distance errors, we obtained
$\sigma_{*,cor}=90.4$ for  $D=1-8$ Mpc, a value which tend to be even lower that
the one derived from observational analysis, 
namely 99.3 km/s (Tikhonov \& Klypin 2009) and with 
a reasonable number of galaxies considered.
For distances from 0.7 to 3 Mpc, we also found a nice agreement between our simulated data and
observational ones from Karachentsev et al. (2009).
Then, these results suggest that there is no particular problem with the \lcdm \ model
in reproducing the velocity dispersion derived from observational analysis.
A similar conclusion was obtained from 
Martinez-Vaquero et
al. (2009) who found 
that a non-negligible fraction of LG like objects simulated form various
cosmological models present
a $\sigma_H$ value close to (and even smaller) than the observed value.
Moreover, similar results have been also obtained from the analysis
of Tikhonov \& Klypin (2009).
Since the present study has been limited to only one realization,
due to the high computational cost when  the physics of baryons is included
in the simulations,
general conclusions cannot be drawn. 
However, one of our main criteria to select LG candidates 
was their relative isolation (no massive galaxies within 3 Mpc).
We stress that we found only one candidate (among 16 MW-M31
 pair candidates) that was satisfying
all the criteria. This raised  the question whether the local group is located in 
a particular place of the universe or not.

To finish, we have also compared our simulated data to observational ones for $R=0.7-3$ Mpc.
The nice agreement between them allows us to test models, in particular the estimation
of the total mass enclosed at the zero-velocity surface radius $R_0$
 using the Lema\^itre-Tolman model adapted to $\Lambda$CDM.
It appears that the estimation of the LG mass using this approach 
is underestimated by at least 50\%. This is  probably
due to the fact that, on the one hand,
 the hypothesis of a spherical infall collapse is not valid anymore
in this specific case. On the other hand, the mass enclosed within  $R_0$
is not constant in the time, as assumed by the TL model (e.g. no mass accretion),
 and this effect may have an important influence in the estimation of the final total mass.

\vspace{1.0cm}

\noindent
{\bf Acknowledgment}

\noindent
I acknowledge support from the ``Agence National de la recherche''
ANR-08-BLAN-0222-02.
It is a pleasure to thank the anonymous referee 
for his/her useful comments which have significantly improved this paper.
I warmly thanks C.\,Alard, S.\,Colombi, J.\,Devriendt,
R.\,Gavazzi, M.\,Hudson, Y.\,Kakazu, I.\,D.\,Karachentsev, A.\,Klypin,
 G.\,Mamon, R.\,Mohayaee, J.\,A. de Freitas Pacheco,
 C.\,Pichon, S.\,Prunet,  J.\,Silk and T.\,Sousbie for interesting discussions.
I also thanks D.\,Munro for
freely distributing his Yorick programming language (available at
\texttt{http://yorick.sourceforge.net/}) which was used during the
course of this work. This work was carried within the framework of the
Horizon project (http://www.projet-horizon.fr).

\end{document}